# The role of nonmagnetic $d^0$ vs. $d^{10}$ *B*-type cations on the magnetic exchange interactions in osmium double perovskites

Hai L. Feng,[a,*] Kazunari Yamaura,[b] Liu Hao Tjeng,[a] Martin Jansen[a,c,*]

[a] Max Planck Institute for Chemical Physics of Solids, Dresden 01187, Germany
[b] Research Center for Functional Materials, National Institute for Materials Science, Tsukuba, Ibaraki 305-0044, Japan
[c] Max Planck Institute for Solid State Research, Stuttgart 70569, Germany

*Corresponding authors
Max Planck Institute for Chemical Physics of Solids, 01187 Dresden, Germany
Hai Feng and Martin Jansen
E-mail: Hai.Feng_nims@hotmail.com (HF) and M.Jansen@fkf.mpg.de (MJ)

**Abstract**

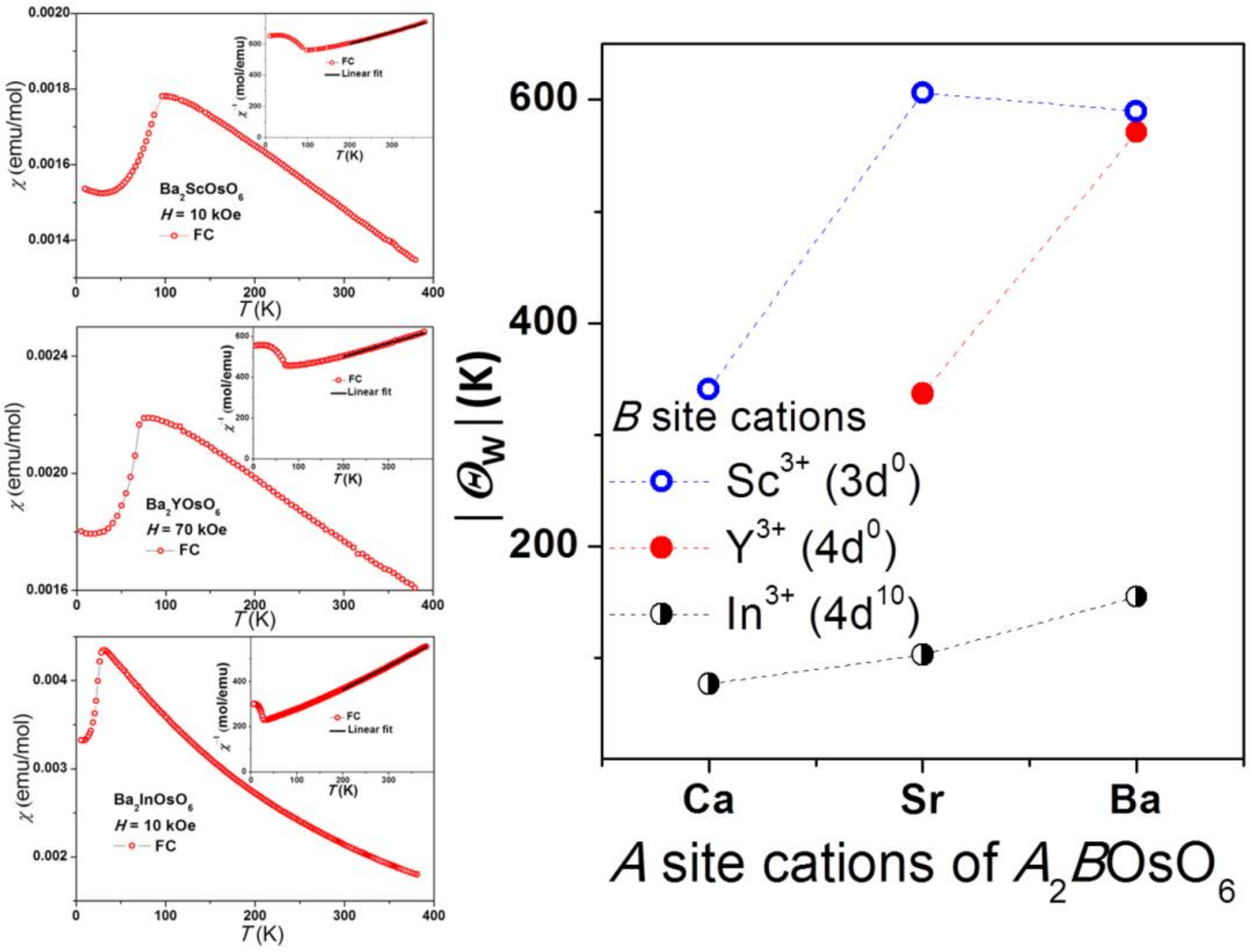

Polycrystalline samples of double perovskites $Ba_2BOsO_6$ ($B$ = Sc, Y, In) were synthesized by solid state reactions. They adopt the cubic double perovskite structures (space group, *Fm*-3*m*) with ordered *B* and Os arrangements. $Ba_2BOsO_6$ ($B$ = Sc, Y, In) show antiferromagnetic transitions at 93 K, 69 K, and 28 K, respectively. The Weiss-temperatures are -590 K for $Ba_2ScOsO_6$, -571 K for $Ba_2YOsO_6$, and -155 K for $Ba_2InOsO_6$. $Sc^{3+}$ and $Y^{3+}$ have the open-shell $d^0$ electronic configuration, while $In^{3+}$ has the closed-shell $d^{10}$. This indicates that a $d^0$ *B*-type cation induces stronger overall magnetic exchange interactions in comparison to a $d^{10}$. Comparison of $Ba_2BOsO_6$ ($B$ = Sc, Y, In) to their Sr and Ca analogues shows that the structural distortions weaken the overall magnetic exchange interactions.

## Introduction

Double perovskite oxides, $A_2BB'O_6$, containing 4d/5d elements have attracted considerable attention due to their remarkable electronic and magnetic properties, such as room-temperature magnetoresistance [1], high-temperature ferrimagnetism [2,3], as well as spin glass [4] and valence bond glass behaviors[5]. As a characteristic structural feature, there are interpenetrating *B* and *B′* face-centered cubic (fcc) sublattices, and their complex magnetic properties are determined by the intricate interplay between the geometrically frustrated intra-sublattice and inter-sublattices exchange interactions, which are correlated with structure distortions [6-11]. The well-studied isoelectronic double perovskites $Sr_2FeOsO_6$ [12-15], and $Ca_2FeOsO_6$ [3,7] are exemplary. $Sr_2FeOsO_6$ crystallizes in a tetragonal double perovskite structure (space group: I4/m) and shows successive antiferromagnetic (AFM) transitions around 140 K and 67 K [12,13], while the isoelectronic $Ca_2FeOsO_6$ crystallizes in a monoclinic variant (space group: $P2_1/n$) and shows room temperature ferrimagnetism [3,7].

When in $A_2BB'O_6$ *B* is a nonmagnetic cation and *B′* is $Os^{5+}$, the magnetic properties are obviously determined by exchange coupling within the $Os^{5+}$ fcc sublattice. Investigations into double perovskites with single $Os^{5+}$ magnetic sublattice will provide help for a better understanding of the remarkable magnetic properties of osmium double perovskites, such as the exceptionally high-temperature ferrimagnetism in $Sr_2CrOsO_6$ [2] and the strikingly different magnetic properties between $Sr_2FeOsO_6$ [12,13] and $Ca_2FeOsO_6$ [3,7]. Most of the well-studied double perovskites with nonmagnetic *B* cations and $Os^{5+}$ forming the *B′* sublattice known so far display AFM transitions [16-23]. Because the magnetic interactions between the nearest-neighbor (NN) $Os^{5+}$ ions are usually thought to run mainly via the $Os^{5+}$-O-O-$Os^{5+}$ path [16,17,20,22], the nonmagnetic *B* cations are not expected to play an important role in the magnetic exchange interactions. However, investigations on $Sr_2BOsO_6$ (*B* = Sc, Y, In) indicate that the electronic configuration of the nonmagnetic *B* cation, $d^0$ or $d^{10}$, does influence the magnetic exchange interactions [16,24]. As $Sr_2BOsO_6$ (*B* = Sc, Y, In) are also structurally distorted, but to a different degree [16], it is better to study the role of nonmagnetic *B* cations in distortion-free analogues. As a simple measure, substituting Ba for Sr would yield cubic double perovskites. In this work, undistorted double perovskites $Ba_2ScOsO_6$, $Ba_2YOsO_6$, and $Ba_2InOsO_6$ were synthesized and their magnetic properties were characterized. By comparing their magnetic properties, we were able to address the role of nonmagnetic $d^0$ vs. $d^{10}$ cations on the magnetic exchange interactions. We have also compared their structural and magnetic properties to those of the Sr and Ca analogues, thereby obtaining information on the effects of structural distortions on the magnetic exchange interactions. The synthesis of $Ba_2ScOsO_6$ and $Ba_2InOsO_6$ was firstly reported by Slight et al. [25]; however, to the best of our knowledge, details of their magnetic properties are not known to date. The synthesis and magnetic properties of $Ba_2YOsO_6$ were published by Kermarrec et al. [23].

## Experimental

Polycrystalline samples of $Ba_2ScOsO_6$ and $Ba_2InOsO_6$ were synthesized by the solid-state reaction from powders of $BaO_2$ (Alfa Aesar, anhydrous, 84% min), $Sc_2O_3$ (Alfa Aesar, 99.9%)/$In_2O_3$ (Alfa Aesar, 99.99%), and $OsO_2$ (Alfa Aesar, 83% Os min). $BaO_2$, $B_2O_3$ ($B$ = Sc or In), and $OsO_2$ were weighted with mole ratio 4:1:2, thoroughly mixed, and pressed into pellets inside a glove box. Then the pellets were transferred into corundum crucibles, which were placed and sealed inside quartz ampules. $Ba_2ScOsO_6$ and $Ba_2InOsO_6$ were obtained after sintering for 96 hours at 1025 ºC and 950 ºC, respectively, with once intermittent regrinding. A polycrystalline sample of $Ba_2YOsO_6$ was synthesized by solid-state reaction from powders of $BaO_2$ (99%, Kojundo Chemical Lab. Co., Ltd.), Os (99.95%, Heraeus Materials Technology), $Y_2O_3$ (99.99%, Kojundo Chemical Lab. Co., Ltd.), and $KClO_4$ (>99.5%, Kishida Chemical Co., Ltd.). The starting materials with stoichiometric ratio were weighted and thoroughly mixed, followed by sealing in a Pt capsule. The sealed Pt capsule was then statically compressed in a belt-type high-pressure apparatus at a pressure of 6 GPa, followed by heating at 1500 °C for 1 hour, while maintaining the high-pressure conditions. It was then quenched to ambient temperature in less than a minute, and subsequently, the pressure was released.

Small pieces of $Ba_2ScOsO_6$ and $Ba_2InOsO_6$ were cut from the synthesized pellets and finely ground for powder X-ray diffraction (Guinier technique, Huber G670 camera. Cu-Kα1 radiation, $\lambda$ = 1.54056 Å, germanium monochromator, 10º ≤ 2θ ≥ 100º, step width of 0.005º). A small piece of $Ba_2YOsO_6$ was finely ground and rinsed in water to remove any KCl residue. The final black powder was investigated by synchrotron X-ray diffraction (SXRD), which was conducted using a large Debye–Scherrer camera installed in the BL15XU beamline in SPring–8. The SXRD data were collected at $\lambda$ = 0.65298 Å (confirmed by a reference material, $CeO_2$) at room temperature. The refinement of crystal structures was carried out by Rietveld analysis using the RIETAN-VENUS software [26,27].

Using pieces of the obtained pellets, the electrical resistivities ($\rho$) were measured with a DC gauge current of 0.1 mA by a four-point method using a physical properties measurement system (PPMS, Quantum Design, Inc.). Electrical contacts were made with Au wires and silver paste. The temperature dependence of the specific heat ($C_p$) was measured in the same PPMS apparatus by using its HC option (relaxation method). The magnetic susceptibilities ($\chi$) of the samples were measured in a SQUID magnetometer (MPMS, Quantum Design), at field cooling (FC) conditions in a temperature range between 2 K and 380 K under applied magnetic fields of 10 kOe for $Ba_2ScOsO_6$ and $Ba_2InOsO_6$, but 70 kOe for $Ba_2YOsO_6$.

**Results and discussions**

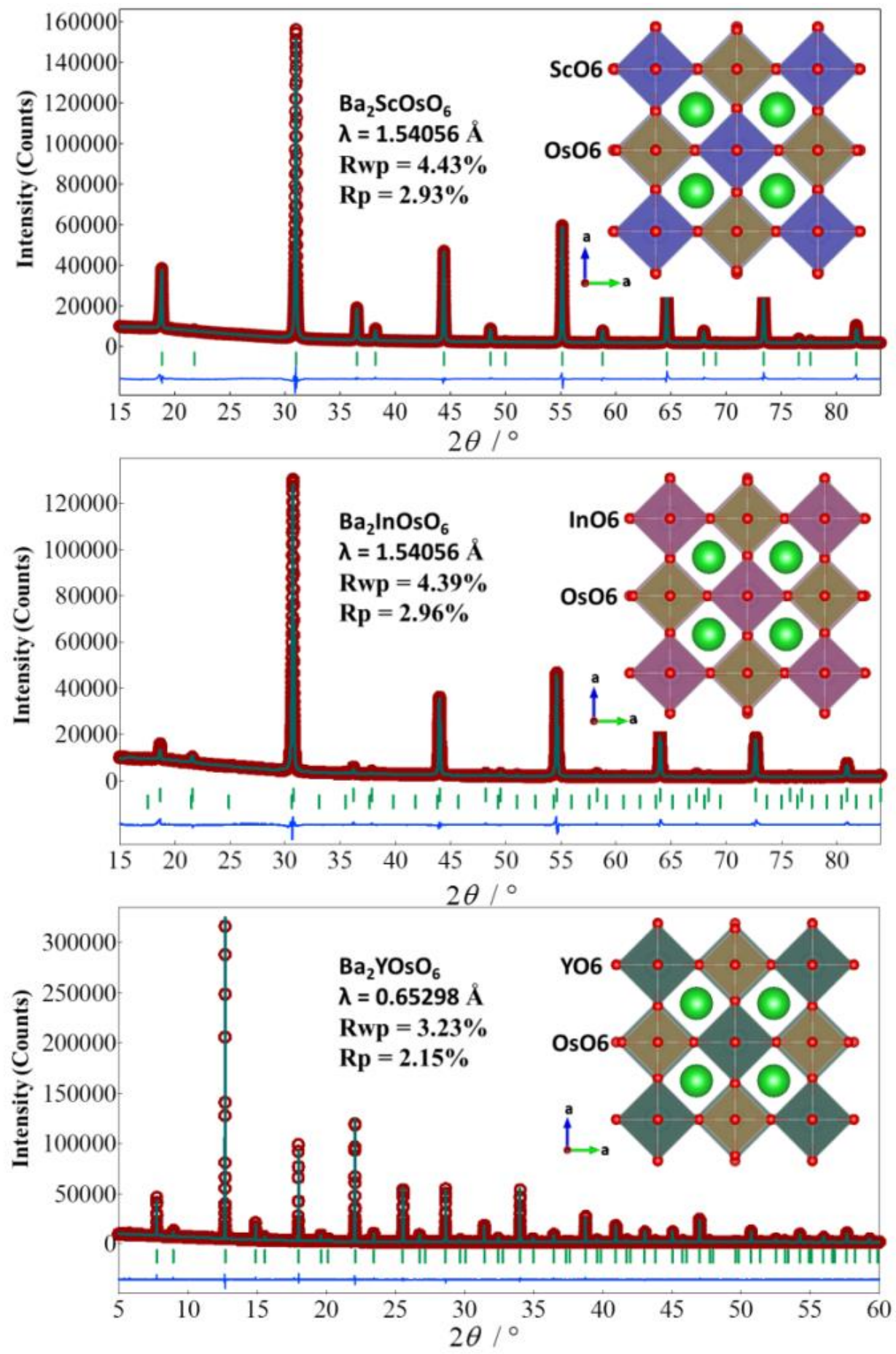

Figure 1. Rietveld refined powder XRD profiles of $Ba_2ScOsO_6$ (top) and $Ba_2InOsO_6$ (middle) and rietveld refined powder SXRD profile of $Ba_2YOsO_6$ (bottom); the insets show the corresponding crystal structures.

The room temperature XRD patterns of $Ba_2ScOsO_6$ and $Ba_2InOsO_6$ and the SXRD pattern of $Ba_2YOsO_6$ are shown in Figure 1. They are well refined assuming cubic double perovskite structures with space group *Fm*-3*m*. The obtained lattice parameters are a = 8.15249(4) Å for $Ba_2ScOsO_6$, a = 8.22424(4) Å for $Ba_2InOsO_6$, and a = 8.35131(1) for $Ba_2YOsO_6$, which are almost as same as those reported, 8.152 Å for $Ba_2ScOsO_6$, 8.224 Å for $Ba_2InOsO_6$, and a = 8.3541(4) for $Ba_2YOsO_6$ [24,25]. The XRD or SXRD pattern of $Ba_2ScOsO_6$ and $Ba_2YOsO_6$, respectively, indicates that the sample is free from impurities. However, in the XRD pattern of $Ba_2InOsO_6$, in addition to the Bragg reflections expected for $Ba_2InOsO_6$, several weak reflections were could be observed, which can be attributed to $In_2O_3$ impurity (about 1% wt). The impurity is diamagnetic and thus does not have any impact on the magnetic properties. During the analysis, the possible anti-site disorder between *B* and Os was carefully checked under the conditions that he 4*a* and 4*b* Wyckoff site are constrained to be fully occupied and the mole ratio *B*/Os were constrained to be 1/1. The occupancies of *B* and Os were then refined with fixed displacement parameters of 0.5. In subsequent cycles the displacement parameters were refined while the occupancies were fixed. This procedure was repeated several times, while the scale factor, peak-shape parameters, asymmetry parameters, lattice parameters and atomic position parameters were refined during the analysis. The obtained atomic positions, occupancy, and isotropic displacement parameters for $Ba_2BOsO_6$ (*B* = Sc, Y, In) at room temperature are shown in Table 1. Within the error limits, the results of the analysis indicated a 100% order of Y and Os, but small amount of disorder of Sc/In and Os. Investigations on crystal structures of ordered double perovskites $A_2M^{3+}TaO_6$ and $A_2M^{3+}NbO_6$ reveal that the compounds showing 100% *B*/*B*′ order usually have large ionic radius difference ($\Delta r \geq 0.26$ Å) between *B* and *B*′ ions [28]. In this work, the Δr for $Ba_2BOsO_6$ (*B* = Sc, Y, In) are 0.17, 0.325, and 0.225 Å, respectively [29], which is consistent with the findings in Ref. 28.

Table 1. Atomic positions, occupancy, and isotropic displacement parameters for $Ba_2BOsO_6$ (*B* = Sc, Y, In) at room temperature.

| Compounds | Atoms | Site | Occupancy | x | y | z | $B_{iso}$ (Å$^2$) |
|---|---|---|---|---|---|---|---|
| $Ba_2ScOsO_6$ | Ba | 8*c* | 1 | 0.25 | 0.25 | 0.25 | 0.28(2) |
| | Sc1/Os1 | 4*a* | 0.98(1)/0.02 | 0 | 0 | 0 | 0.20(2) |
| | Os2/Sc2 | 4*b* | 0.98/0.02 | 0.5 | 0 | 0 | 0.19(2) |
| | O | 24*e* | 1 | 0.2618(3) | 0 | 0 | 0.16(6) |
| $Ba_2YOsO_6$ | Ba | 8*c* | 1 | 0.25 | 0.25 | 0.25 | 0.50(1) |
| | Y | 4*a* | 1 | 0 | 0 | 0 | 0.24(2) |
| | Os | 4*b* | 1 | 0.5 | 0 | 0 | 0.26(1) |
| | O | 24*e* | 1 | 0.2669(3) | 0 | 0 | 0.49(5) |
| $Ba_2InOsO_6$ | Ba | 8*c* | 1 | 0.25 | 0.25 | 0.25 | 0.24(3) |
| | In1/Os1 | 4*a* | 0.97(1)/0.03 | 0 | 0 | 0 | 0.43(4) |
| | Os2/In2 | 4*b* | 0.97/0.03 | 0.5 | 0 | 0 | 0.43(3) |
| | O | 24*e* | 1 | 0.2624(7) | 0 | 0 | 0.34(8) |

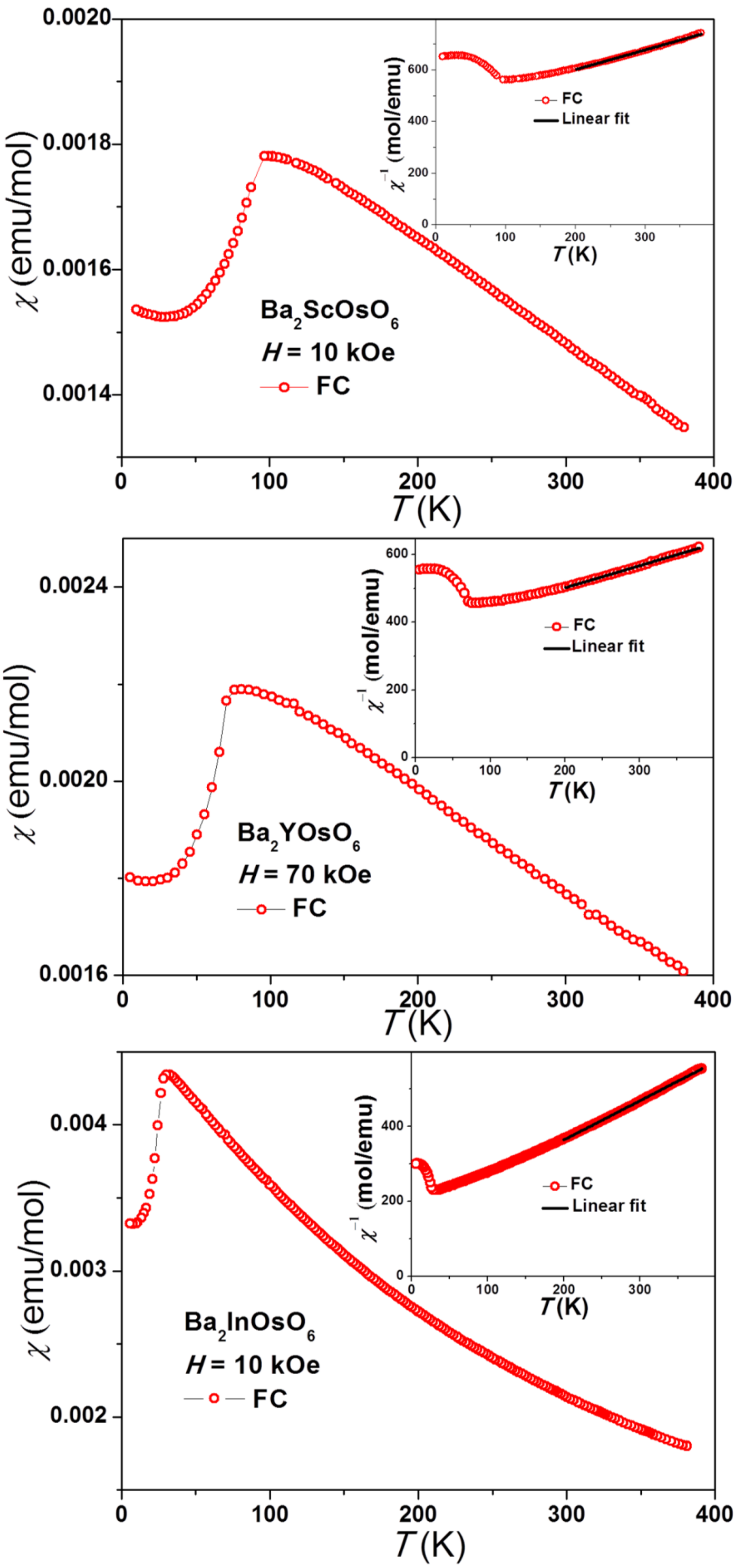

Figure 2. Temperature dependent magnetic susceptibilities of $Ba_2ScOsO_6$ (top), $Ba_2YOsO_6$ (middle), and $Ba_2InOsO_6$ (bottom). Insets show the corresponding plots of their inverse susceptibilities.

The temperature dependent magnetic susceptibilities, $\chi(T)$, of $Ba_2ScOsO_6$, $Ba_2YOsO_6$, and $Ba_2InOsO_6$ are shown in Figure 2. $Ba_2ScOsO_6$, $Ba_2YOsO_6$, and $Ba_2InOsO_6$ show a maximum at temperatures of 93 K, 69 K, and 28 K, respectively, indicating the onset of AFM ordering. In comparision to $Ba_2InOsO_6$, the curvature of $\chi(T)$ for $Ba_2ScOsO_6$ and $Ba_2YOsO_6$ is concave upon approach to Néel temperature ($T_N$), suggesting the presence of short range magnetic correlations. To further characterize these magnetic transitions, specific heat was measured (see Figure 3). Lambda-type anomalies can be observed around temperatures of 93 K, 69 K, and 28 K for $Ba_2ScOsO_6$, $Ba_2YOsO_6$, and $Ba_2InOsO_6$, respectively, confirming the formation of long range AFM order.

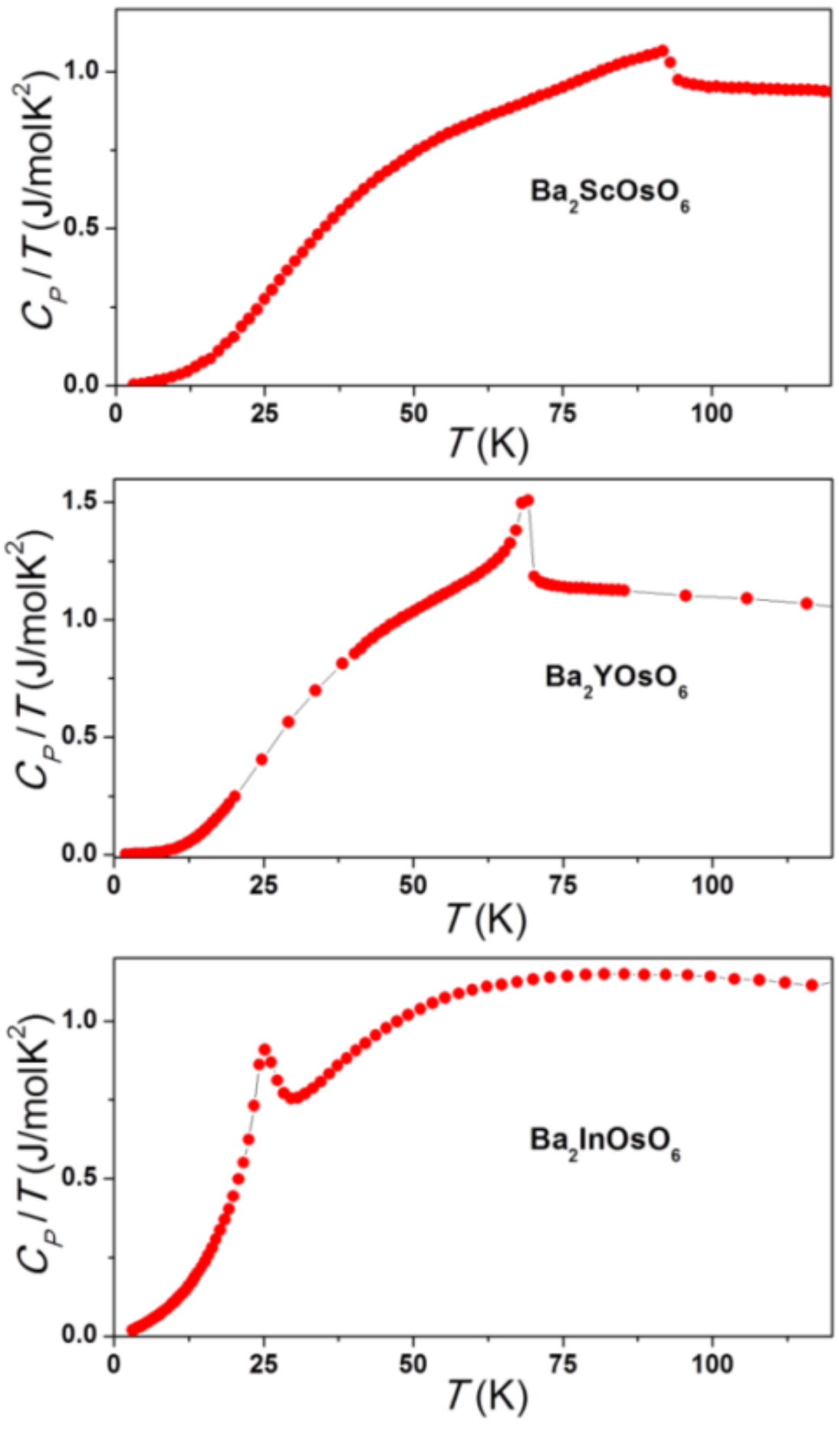

Figure 3. Temperature dependence of the specific heat of $Ba_2ScOsO_6$ (top), $Ba_2YOsO_6$ (middle), and $Ba_2InOsO_6$ (bottom).

Curie-Weiss analysis of the high-temperature part of the inverse susceptibility has resulted in the effective moment, $\mu_{eff}$, of 3.26 $\mu_B$ for $Ba_2ScOsO_6$, 3.45 $\mu_B$ for $Ba_2YOsO_6$, and 2.89 $\mu_B$ for $Ba_2InOsO_6$. The extracted Weiss-temperatures, $\Theta_W$, are -590 K for $Ba_2ScOsO_6$, -538 K for $Ba_2YOsO_6$, and -181 K for $Ba_2InOsO_6$. These $\mu_{eff}$ values are comparable with most of the values reported for $Os^{5+}$, in the range of 2.71-3.72 $\mu_B$ [16-22,30], but they are lower than the theoretical spin-only moment (3.87 $\mu_B$) for $Os^{5+}$ ($S$ = 3/2), which can be partially attributed to the reduction of $Os^{5+}$ $\mu_{eff}$ due to spin-orbit coupling. It has been suggested that the effect of spin-orbit coupling cannot be ignored in $4d^3$ and $5d^3$ systems [31].

The negative $\Theta_W$ reflects that AFM correlations are predominant in these compounds. $Ba_2ScOsO_6$ and $Ba_2InOsO_6$ are electrically semiconducting (see Figure 4), similar to other double perovskites with single $Os^{5+}$ magnetic sublattices [16-22]. Attempts were made to plot the data on a $T^{-1}$ and $T^{-1/4}$ scales, and both samples are found to be linear on a $T^{-1/4}$ scale as shown in the inset of Figure 4, in accordance with a three-dimensional variable range hopping transport model. Their magnetic exchange interactions should be dictated by extended superexchange interactions such as $Os^{5+}$-O-O-$Os^{5+}$. The large negative $\Theta_W$ of -590 K for $Ba_2ScOsO_6$ and -538 K for $Ba_2YOsO_6$ indicate very strong AFM exchange interactions. Strikingly, the $\Theta_W$ for $Ba_2InOsO_6$ is only -181 K. Because they all crystallize in the same cubic double perovskite structure, the difference among $Ba_2BOsO_6$ ($B$ = Sc, Y, In) is that they have different nonmagnetic $B$-type cations. $Sc^{3+}$ and $Y^{3+}$ have the open-shell $3d^0$ and $4d^0$ electronic configurations, respectively, but $In^{3+}$ has the closed-shell $4d^{10}$ electronic configuration. The large change of $\Theta_W$ from -538 K and -590 K for $Ba_2ScOsO_6$ and $Ba_2YOsO_6$, respectively, to -181 K for $Ba_2InOsO_6$ indicates that a $d^0$ $B$-type cation induces stronger overall magnetic exchange interactions in comparison to a $d^{10}$ $B$-type cation. Here we note that in our undistorted cubic double perovskites, an s or p orbital at the $B$-type cation cannot mediate the NN $Os^{5+}$-O-$B$-O-$Os^{5+}$ exchange interaction ($Os^{5+}$-$t_{2g}^3$, $Os^{5+}$-$B$-$Os^{5+}$-90º bond angle) so that the difference between the Sc/Y vs the In compounds must be attributed to the difference in the d-configuration. Indeed, the first-principles density functional theory studies of $Sr_2BOsO_6$ ($B$ = Sc, Y, In) reveal that the electronic configuration of nonmagnetic $B$ cations, either open shell $d^0$ or closed shell $d^{10}$, has an increasing or decreasing influence on the hybridization between the Os-5d and Sc/Y/In-d states, and it was found that the exchange coupling depends strongly on the overlap between Os-5d and Sc/Y/In-d states [24]. The hybridization between Os-5d and Sc/Y/In-d states is much smaller in the $d^{10}$ closed-shell case than in the $d^0$ open-shell cases, which results in the smaller amplitudes of Os-Os coupling in the $In^{3+}$ compound than in the $Y^{3+}$ and $Sc^{3+}$ compounds [24].

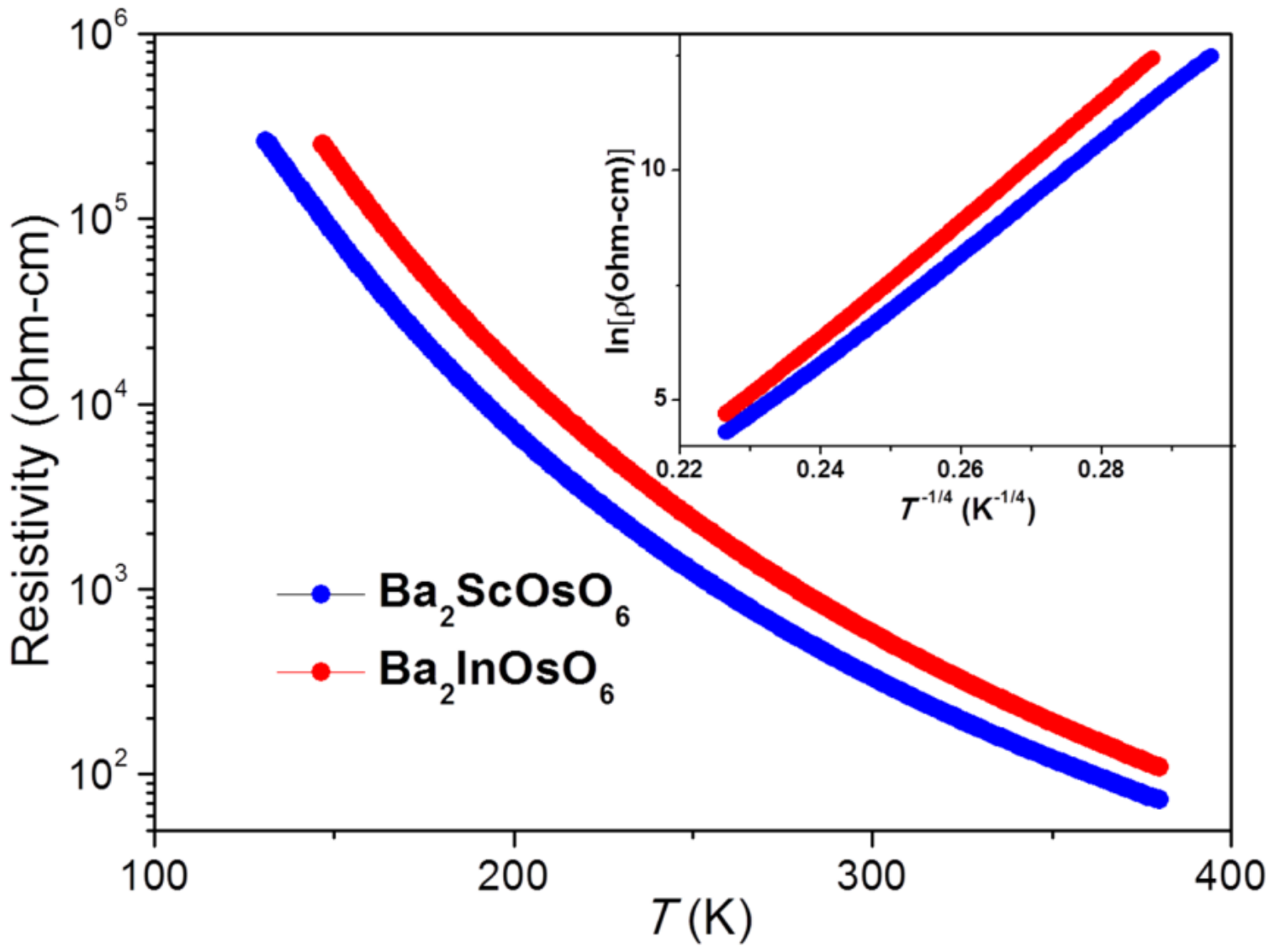

Figure 4. Electrical resistivity of $Ba_2ScOsO_6$ and $Ba_2InOsO_6$

In double perovskites $Ba_2BOsO_6$ ($B$ = Sc, Y, In), $Os^{5+}$ ions form fcc sublattices, and they are magnetically frustrated when the AFM NN $Os^{5+}$-$Os^{5+}$ interactions are predominant. The $T_N$s of $Ba_2BOsO_6$ ($B$ = Sc, Y, In) are 93 K, 69 K, and 28 K, respectively, which are much lower than their $|\Theta_W|$ values, clearly indicating the presence of magnetic frustration. The frustration factor, $|\Theta_W/T_N|$, is 6.3, 7.8, and 6.5 for $Ba_2ScOsO_6$, $Ba_2YOsO_6$, and $Ba_2InOsO_6$, respectively. $Ba_2YOsO_6$ was reported to adopt the type-I fcc spin structure [23]. In type-I spin structure, each osmium is antiferromagnetically coupled to eight out of twelve NN ions, but the NNN ions are all ferromagntically coupled. It is usually stabilized when the AFM NN interactions are predominant, whereas the next-nearest-neighbor interactions are either ferromagnetic or negligibly small [32]. Because $Ba_2BOsO_6$ ($B$ = Sc, Y, In) are magnetic frustrated and have comparable values of $|\Theta_W/T_N|$, $Ba_2ScOsO_6$ and $Ba_2InOsO_6$ may adopt the same type-I spin structures as $Ba_2YOsO_6$. The similar compounds $Sr_2BOsO_6$ ($B$ = Sc, Y, In) were reported to adopt type-I spin structures [16,17]. Further neutron diffraction and theoretical calculation are needed for the determination of magnetic structures of $Ba_2ScOsO_6$ and $Ba_2InOsO_6$.

The space groups, lattice parameters, average bond angles of $B$-O-Os, and magnetic properties of double perovskites $Ba_2BOsO_6$ ($B$ = Sc, Y, In) and their Sr and Ca analogues are summarized in Table 2. $Ba_2BOsO_6$ ($B$ = Sc, Y, In) crystallize in the ideal double perovskite structures, in which there is no structural distortion and the $B$-O-Os bond angles are 180°. When $Ba^{2+}$ is replaced by the smaller cation $Sr^{2+}$, the crystal structures distort and the *B*-O-Os bond angles deviate from 180º in order to accommodate the smaller *A* site cation. When $Sr^{2+}$ is replaced by the even smaller cation $Ca^{2+}$, the crystal structures are even more distorted with more buckled *B*-O-Os bonds.

Generally, the values of $|\Theta_W|$ for $Ba_2BOsO_6$ (*B* = Sc, Y, In) are larger than those of their Sr and Ca analogues, suggesting that the strength of the overall $Os^{5+}$-$Os^{5+}$ magnetic exchange interactions correlates negatively with structural distortions: the structural distortions weaken the overall magnetic exchange interactions. All these double perovskites with single $Os^{5+}$ magnetic sublattice display long range AFM order and show magnetic frustration as indicated by an average $|\Theta_W/T_N|$ value of ≈6.5. Apparently, there is no clear correlation between the $|\Theta_W/T_N|$ and the structural distortions.

Table 2. Double perovskites with single $Os^{5+}$ magnetic sublattice.

| Compound $A_2BOs^{5+}O_6$ | Space group | Lattice parameters (Å or ° or Å$^3$) | Average ∠$B$-O-Os (°) | $T_N$ (K) | $\Theta_W$ (K) | $\mu_{eff}$ ($\mu_B$) | $\lvert\Theta_W/T_N\rvert$ | Ref. |
|---|---|---|---|---|---|---|---|---|
| $Ca_2ScOsO_6$ | $P2_1/n$ | a = 5.4716(1)<br>b = 5.6165(1)<br>c = 7.8168(1)<br>β = 89.889(2)<br>V/Z = 120.109 | 151.3 | 69 | -341 | 3.72 | 4.9 | [22] |
| $Sr_2ScOsO_6$ | $P2_1/n$ | a = 5.6666(1)<br>b = 5.6428(1)<br>c = 7.9791(1)<br>β = 90.083(2)<br>V/Z = 127.620 | 166.1 | 92 | -606 | 2.99 | 6.5 | [16] |
| $Ba_2ScOsO_6$ | $Fm$-3$m$ | a = 8.15249(4)<br>V/Z = 135.460 | 180 | 93 | -590 | 3.26 | 6.3 | this work |
| $Sr_2YOsO_6$ | $P2_1/n$ | a = 5.7817(1)<br>b = 5.8018(1)<br>c = 8.1877(1)<br>β = 90.227(1)<br>V/Z = 136.945 | 157.3 | 53 | -337 | 3.45 | 6.4 | [16] |
| $Ba_2YOsO_6$ | $Fm$-3$m$ | a = 8.3541(4)<br>V = 145.760 | 180 | 67,69 | -717 | 3.93 | 10.4 | [23] |
| | | a = 8.35131(1)<br>V/Z = 145.614 | 180 | 69 | -571 | 3.52 | 8.3 | this work |
| $Ca_2InOsO_6$ | $P2_1/n$ | a = 5.4889(3)<br>b = 5.6785(1)<br>c = 7.8576(2)<br>β = 90.12(2)<br>V/Z = 122.456 | 147.4 | 14 | -77 | 3.15 | 5.5 | [18] |
| $Sr_2InOsO_6$ | $P2_1/n$ | a = 5.7002(1)<br>b = 5.6935(1)<br>c = 8.0521(1)<br>β = 90.112(1)<br>V/Z = 130.695 | 158.5 | 26 | -103 | 3.08 | 4.0 | [16] |
| $Ba_2InOsO_6$ | $Fm$-3$m$ | a = 8.22424(4)<br>V/Z = 139.068 | 180 | 28 | -155 | 2.84 | 5.5 | this work |

## Conclusions

Polycrystalline samples of $Ba_2BOsO_6$ (*B* = Sc, Y, In) were synthesized by solid state reactions. Their structural and magnetic properties were investigated. Room temperature XRD showed that they all crystallize as ordered double perovskites with space group *Fm*-3*m*. Magnetic susceptibilities and specific heat measurements revealed that $Ba_2BOsO_6$ (*B* = Sc, Y, In) order antiferromagnetically at 93 K, 69 K, and 28 K, respectively. The extracted $\Theta_W$ is -590 K for $Ba_2ScOsO_6$, -571 K for $Ba_2YOsO_6$, and -155 K for $Ba_2InOsO_6$. $Sc^{3+}$ and $Y^{3+}$ have the open-shell $d^0$ electronic configurations, but $In^{3+}$ has the closed-shell $d^{10}$ electronic configuration, indicating that a $d^0$ *B*-type cation induces stronger overall AFM exchange interactions in comparison to a $d^{10}$ *B*-type cation. Comparison of $Ba_2BOsO_6$ (*B* = Sc, Y, In) to the Sr and Ca analogues showed that the structural distortions weaken the overall AFM exchange interactions.

## Acknowledgments

The research in Dresden was partially supported by the Deutsche Forschungsgemeinschaft through SFB 1143. The research in NIMS was supported in part by JSPS KAKENHI Grant Numbers JP16H04501 and JP15K14133.